\documentclass[pre,aps,twocolumn,amsmath,amssymb,longbibliography]{revtex4-1}

\pdfoutput=1
\usepackage{hyperref}
\usepackage{graphicx}
\usepackage[usenames]{color}
\usepackage{bm}
\usepackage[final]{movie15}

\definecolor{blue}{rgb}{0,0,0}

\def\cal#1{\mathcal{#1}}
\def\eqq#1{Eq.~(\ref{#1})}
\def\eq#1{(\ref{#1})}
\def\av#1{\langle #1 \rangle}

\def\f#1{Fig.~\ref{#1}}

\def\c#1{~\cite{#1}}
\def\cc#1{Ref.\c{#1}}

\def\pp{{\rm Pe}}

\def\ta0{\tilde{a}_0}

\def\s#1{Section~\ref{#1}}
\def\kt{k_{\rm B}T}
\def\e{{\rm e}}

\def\beq{\begin{equation}}
\def\eeq{\end{equation}}
\def\bea{\begin{eqnarray}}
\def\eea{\end{eqnarray}}

\begin{document}

\title{Low-dissipation self-assembly protocols of active sticky particles}
\author{Stephen Whitelam$^1$}
\email[]{swhitelam@lbl.gov}
\author{Jeremy D. Schmit$^2$}
\email[]{schmit@phys.ksu.edu}
\affiliation{$^1$Molecular Foundry, Lawrence Berkeley National Laboratory, 1 Cyclotron Road, Berkeley, CA 94720, USA\\
$^2$Department of Physics, Kansas State University, Manhattan, KS, 66506, USA}

\begin{abstract}
We use neuroevolutionary learning to identify time-dependent protocols for low-dissipation self-assembly in a model of generic active particles with interactions. When the time allotted for assembly is sufficiently long, low-dissipation protocols use only interparticle attractions, producing an amount of entropy that scales as the number of particles. When time is too short to allow assembly to proceed via diffusive motion, low-dissipation assembly protocols instead require particle self-propulsion, producing an amount of entropy that scales with the number of particles and the swim length required to cause assembly. Self-propulsion therefore provides an expensive but necessary mechanism for inducing assembly when time is of the essence.
\end{abstract}

\maketitle
\section{Introduction}

Biological processes depend on the assembly of molecular complexes in the face of many constraints\c{Phillips2008}. Molecular self-assembly is most effective when it occurs in a short time, with reproducible high yield, and exerts minimal metabolic demands on the cell, but often these constraints are not mutually reconcilable. For instance, the most energy-efficient mode of self-assembly is to allow molecular complexes to find each other through diffusion, but diffusion is slow, requiring timescales that grow as the square of characteristic interparticle separations. Thus while a nanometer protein can diffuse across a micron-sized prokaryotic cell in milliseconds, eukaryotic
cells, which have sizes on the order of 10 microns, must be sub-divided by membrane-bound organelles so that diffusion within them can be efficient. When such subdivision is not possible, such as for transport of neurotransmitters along neuronal axons, cells use active transport by molecular motors to obtain acceptable
transit times. However, motors require ATP as fuel and so consume energy\c{Phillips2008}.

In this paper we explore the competing requirements of speed and energy efficiency within a statistical mechanical model of self-assembly. Motivated by the example of intracellular self-assembly we consider a model that represents in a schematic way some of the key features of intracellular complexes, but is not intended to represent a specific biological example. In the model, complexes are represented as particles on a lattice. Particles can diffuse and bind to each other. They can also be active, able to move persistently in a particular direction\c{needleman2017active,ramaswamy2010mechanics,fodor2018statistical,hagan2016emergent}. We set them the task of self-assembling in a fixed time, and use evolutionary methods to identify the time-dependent protocols associated with directional motion and binding that achieve this goal with as little entropy production as possible. A corresponding physical scenario relates to the control of activity via the modulation of ATP, combined with the modulation of binding affinities. Examples of the latter include the GTP-dependent stabilization of microtubules\c{Desai1997}, turnover of active chaperones\c{Thirumalai2020}, and the cyclical progression of
molecular motors\c{Gennerich2009}. We choose entropy production as a convenient measure of the distance from equilibrium of a dynamical trajectory\c{schnakenberg1976network,ruelle1997entropy}, without concern for whether or not its minimization is an organizing principle of biology. 

We find that the nature of low-entropy production (or low-dissipation) self-assembly protocols within the model depends on the time allowed for assembly. When there is sufficient time for diffusion to achieve mass transport on the required scales, assembly relies on diffusion and energetic interactions only. However, when the time allowed for assembly drops below a certain threshold, the combination of diffusion and energetic binding cannot achieve assembly. In this case, particles must undergo self-propulsion, which comes with a large energetic cost. These results, interpreted in the context of cellular assembly, suggest an incentive to maintain the system size small enough for diffusion-driven assembly to occur on the required timescales.

In more detail, we encode the time-dependent protocol of the model using a neural network, and use neuroevolutionary methods\c{GA,GA2,floreano2008neuroevolution,salimans2017evolution,Guber,whitelam2020learning,whitelam2021neuroevolutionary} to learn the protocols that achieve a specified degree of self-assembly with the least entropy production possible. There is no guarantee that a given learning scheme will identify the optimum protocol for a given objective, which even in simple systems can involve abrupt changes of control parameters\c{schmiedl2007optimal,solon2018phase,ye2022optimal} (although neural networks of sufficient size can express arbitrary smooth functions, including rapidly-changing ones, and so constitute a good starting point for such a search). To test the learning algorithm we make contact with previous work by learning protocols for low-dissipation magnetization reversal in the Ising model; the protocols learned by the present method are consistent with those learned by path-sampling methods\c{rotskoff2015optimal,gingrich2016near}, and are essentially equivalent to the least-dissipation pathways for that model. For the active model on which we focus we find low-dissipation pathways of two kinds, involving self-propulsion or diffusion alone, according to how much time is allotted for assembly. The two mechanisms produce amounts of entropy that differ considerably. We argue that this competition should be present generically, suggesting a principle that natural systems must observe in order to effect a change of phase with least energy expenditure. 

In \s{isingsec} we test the evolutionary framework by making contact with previous work. In \s{beasties} we apply it to a model of active particles with interactions. We conclude in \s{conc}.

\section{Low-dissipation magnetization reversal protocols in the Ising model}
\label{isingsec}

We start with the 2D Ising model on a square lattice\c{onsager1944crystal,binney1992theory}. The lattice has $N=50^2$ sites, with periodic boundary conditions in both directions. On each site $i$ is a binary spin $S_i = \pm 1$, and the lattice possesses an energy function
\beq
\label{ising}
E = -J\sum_{\av{ij}} S_i S_j -h\sum_{i=1}^N S_i.
\eeq
Here $J$ (which we set to 1) is the Ising coupling and $h$ is the magnetic field. The first sum in \eq{ising} runs over all nearest-neighbor bonds, while the second runs over all lattice sites. We begin with all spins up, giving magnetization $m=N^{-1} \sum_{i=1}^N S_i=1$, and carry out a Glauber Monte Carlo simulation. At each step of the algorithm a lattice site $i$ is chosen at random, and a proposed change $S_i \to -S_i$ made. The change is accepted with probability
\beq
\label{glauber}
p(\Delta E) = \left( 1+ \exp(\beta \Delta E) \right)^{-1},
\eeq
where $\Delta E$ is the energy change under the proposed move, and is $\beta$ the reciprocal temperature in units such that $k_{\rm B} = 1$. If the move is rejected, the original spin state is adopted. We start with $h=1$ and $\beta=1$, and so are below the Ising model critical temperature, $\beta > \beta_{\rm c} =(2J)^{-1}\ln\left(1 + \sqrt{2} \right)  \approx 0.44$\c{onsager1944crystal}, in the two-phase region of the phase diagram.
\begin{figure}[] 
   \centering
  \includegraphics[width=\linewidth]{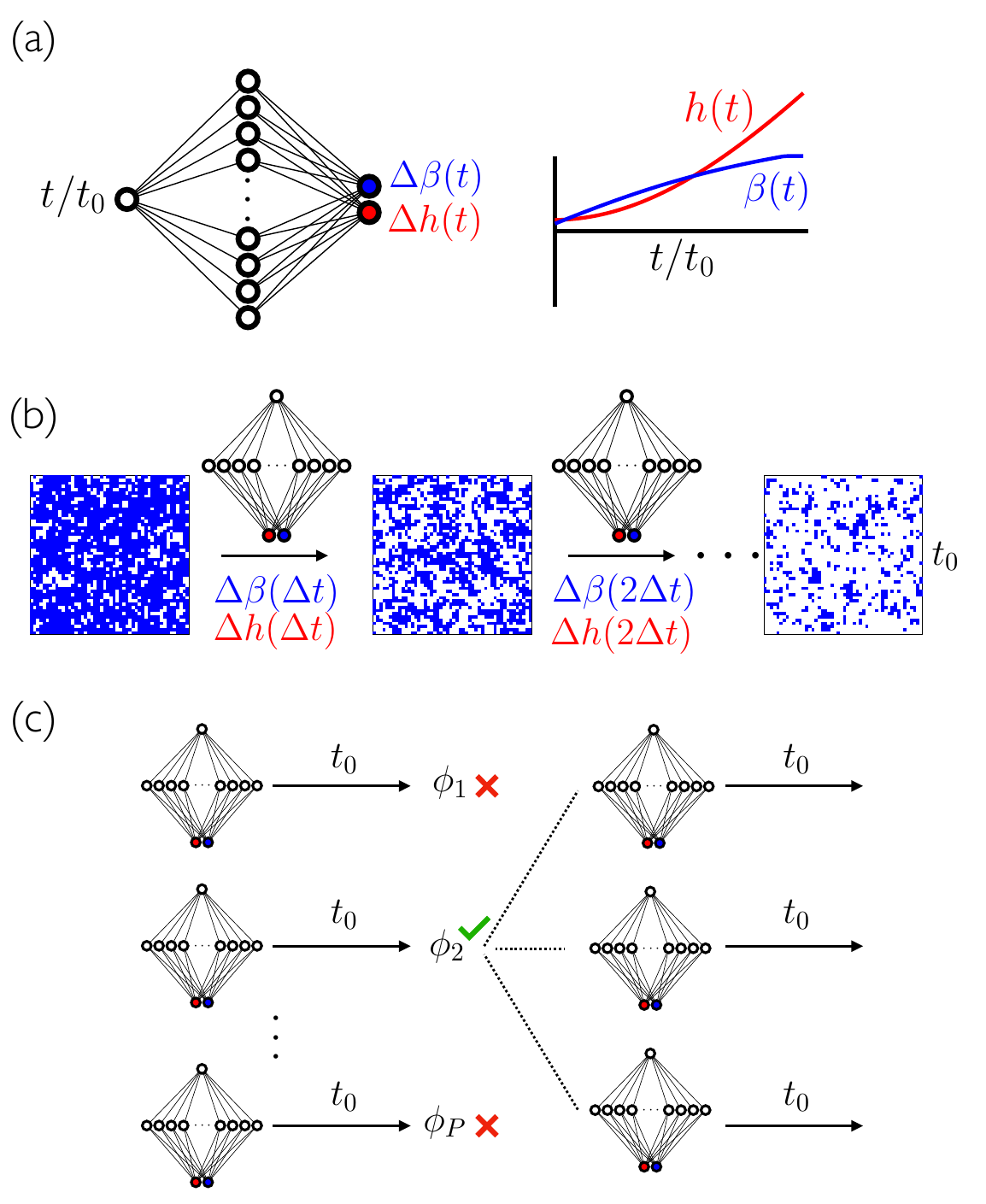} 
   \caption{Schematic of the neuroevolutionary learning method. (a) A time-dependent protocol, which in \s{isingsec} comprises $(\beta(t),h(t))$, is encoded by a neural network whose input is scaled time $t/t_0$ and whose output is the vector $(\Delta \beta(t), \Delta h(t))$. (b) The protocol is imposed within a molecular simulation by the neural network, which acts at 1000 evenly-spaced time increments $\Delta t=t_0/1000$. (c) An evolutionary algorithm searches for the neural network (and hence protocol) that maximizes a particular order parameter, $\phi$, \eqq{phi1}. A population of $P$ randomly-initialized networks, each controlling one molecular simulation, results in $P$ outcomes $\phi_i$, $i\in[1,P]$. In this schematic the second outcome (``individual'') is the best, $\phi_2 >\phi_{j\neq 2}$, and so that neural network is cloned and mutated in order to build the second generation of the evolutionary scheme. In \s{isingsec} we set $P=100$, and the top 10 individuals are retained from one generation to the next.}
   \label{fig0}
\end{figure}

\begin{figure*}[] 
   \centering
  \includegraphics[width=0.8\linewidth]{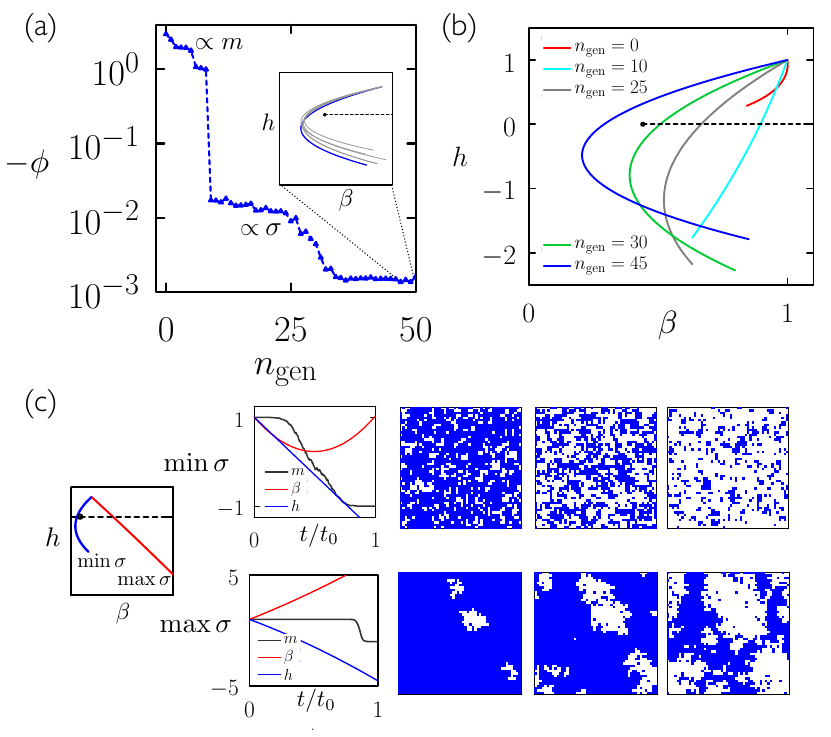} 
   \caption{Evolutionary learning identifies low-dissipation magnetization-reversal protocols in the Ising model. (a) Order parameter \eq{phi1} as a function of evolutionary time (inset: parametric protocols for generations 46--50). (b) Parametric protocols $(\beta(t),h(t))$ for 5 different generations; the black dot and dashed line are the Ising model critical point and first-order phase transition line, respectively. (c) Contrasting low- and high-dissipation magnetization-reversal protocols (from left to right: parametric protocols; time-dependent protocols; time-ordered snapshots).}
   \label{fig1}
\end{figure*}

Our goal is to find a time-dependent protocol $(\beta(t), h(t))$ that will reverse the magnetization of the model with as little entropy produced as possible (i.e. as can be found by the search algorithm). To do so we encode the protocol using a fully connected single-layer neural network, as shown in \f{fig0}(a). The network has one input neuron, which takes the value $t/t_0 \leq 1$, where $t$ is the current number of Monte Carlo steps performed and $t_0=100 N$ the length of the simulation. The neural network has $K=1000$ hidden neurons, with hyperbolic tangent activation functions, and two output neurons, which return the values $(\Delta \beta_{\bm \theta}(t),\Delta h_{\bm \theta}(t))$, where
\beq
\label{nn1}
\Delta \beta_{\bm \theta}(t)=K^{-1/2} \sum_{i=1}^{K-1} \theta_{3i} \tanh\left( \theta_{3 i+1} t/t_0+ \theta_{3 i+2}\right),
\eeq
and
\beq
\label{nn2}
\Delta h_{\bm \theta}(t)=K^{-1/2} \sum_{i=K}^{2K-1} \theta_{3i} \tanh\left( \theta_{3 i+1} t/t_0+ \theta_{3 i+2}\right).
\eeq
Equations~\eq{nn1} and~\eq{nn2} specify the neural network. Here ${\bm \theta}=(\theta_1,\dots,\theta_i,\dots,\theta_{6K})$ is a 6$K$-dimensional vector of parameters (weights and biases), where $K$ is the number of hidden neurons. Eqns.~\eq{nn1} and~\eq{nn2} can, for large enough $K$, approximate any smooth functions $\Delta \beta(t)$ and $\Delta h(t)$\c{cybenko1989approximation}. Our goal will be to adjust the parameters ${\bm \theta}$ until the functions $\Delta \beta_{\bm \theta}(t)$ and $\Delta h_{\bm \theta}(t)$ expressed by the neural network achieve our goal of magnetization reversal with as little entropy produced as possible.

 The neural network acts at 1000 evenly-spaced time increments, as sketched in \f{fig0}(b). Each time it acts, the control parameters are adjusted as
\bea
\beta(t) &\to& \max(0,\beta(t)+\Delta \beta_{\bm \theta}(t)), \, {\rm and} \\
h(t) &\to& h(t)+\Delta h_{\bm \theta}(t).
\eea
We quantify the outcome of the simulation using the order parameter
\beq
\label{phi1}
\phi \equiv \begin{cases}
-m(t_0)-c_1 \quad (m(t_0)>-1) \\
-c_2 \sigma(t_0) \qquad \,\,\,(m(t_0)=-1)
\end{cases}.
\eeq
Here $m(t_0)$ is magnetization at the end of the simulation, $c_1=2$ and $c_2=1/t_0$ are constants, and $\sigma(t_0)$ is the entropy produced during the simulation. There are a number of ways of defining entropy production for a stochastic trajectory\c{ruelle1997entropy}; here we consider the entropy change of the medium\c{schnakenberg1976network,seifert2005entropy}, a sum of terms $\ln(w/w')$ for each process that occurs in the trajectory, where $w$ is the rate for the process and $w'$ the rate for its reverse. In equilibrium, the mean value of this sum is zero.

For the Ising model, the entropy production of the medium for each move is
\beq
\label{entprod1}
\Delta \sigma = \ln \frac{p(\Delta E)}{p(-\Delta E)}=-\beta \Delta E,
\eeq
using \eq{glauber}, and we define $\sigma(t_0)$ as the sum of terms \eq{entprod1} for each move made (note that both $\beta$ and the value of $h$ appearing in $\Delta E$ are time dependent). The order parameter \eq{phi1} imposes two objectives: it is maximal for a simulation whose final value of magnetization is $-1$, and for which as little entropy as possible was produced. The constants $c_1$ and $c_2$ ensure that the value of $\phi$ for any simulation for which $m(t_0)=-1$ exceeds that for which $m(t_0)\neq -1$, and the former simulations are then distinguished by their entropy produced.

To determine a protocol that maximizes \eq{phi1}, i.e. that results in magnetization reversal with as little entropy produced as possible, we use evolutionary learning on the parameters ${\bm \theta}$ of the protocol-encoding neural networks (sometimes called neuroevolution)\c{GA,GA2,floreano2008neuroevolution,salimans2017evolution,Guber,whitelam2020learning,whitelam2021neuroevolutionary}. This protocol is sketched in \f{fig0}(c). We begin with Generation 0, a population of $P=100$ independent simulations each controlled by a distinct neural network. Each neural network's parameters $\theta_i$ are independently and randomly initialized as Gaussian random numbers with zero mean and variance $\delta_1^2$, i.e. $\theta_i \sim {\cal N}(0,\delta_1^2)$. The variance $\delta_1^2$ is also a random number, with $\delta_1 \sim |{\cal N}(0,\delta_0^2)|$, drawn once for each neural network. The parameter $\delta_0=10^{-3}$. Thus each network enacts a different, random protocol. Generation 0 is run for Monte Carlo time $t_0$, and the 10 neural networks resulting in the 10 simulations with the largest values of $\phi$, \eqq{phi1}, are chosen to be the parents of Generation 1. To construct the 100 members of Generation 1, we pick 100 times randomly with replacement from the set of 10 parents, and mutate the chosen neural networks by the addition of Gaussian random numbers 
\beq
\theta_i \to \theta_i+{\cal N}(0,\delta_1^2)
\eeq to each of their parameters. The 100 simulations of Generation 1 are run for Monte Carlo time $t_0$, ranked by their values of $\phi$, and the 10 highest-ranking neural networks become the parents of Generation 2. As this process continues over evolutionary time, the features of the time-dependent protocols that result in increasingly large values of $\phi$ are passed to subsequent generations and refined.

In \f{fig1} we show the outcome of this evolutionary learning procedure. Panel (a) shows $-\phi$ for the highest-ranking protocol of each generation as a function of the number $n_{\rm gen}$ of generations of evolutionary learning. After about 10 generations the best protocol has succeeded in reversing the magnetization of the Ising model, and $\phi$ jumps between the first and second clauses in \eq{phi1}. Thereafter, protocols evolve so as to minimize entropy production while achieving magnetization reversal.

In \f{fig1}(b) we show parametric plots of protocols ($\beta(t), h(t))$ for the best protocols from 5 different generations. Protocols start at the point $(1,1)$. The Ising model (infinite system size) critical point and first-order phase transition line are shown as a black dot and black dashed line, respectively. Comparison with panel (a) shows that the first protocols that achieve magnetization reversal pass through the phase transition line. As their entropy production is progressively reduced, protocols move toward the critical point and then pass around it, avoiding the phase transition line and the vicinity of the critical point. To provide context for the scale of \f{fig1}(a), note that the entropy produced by flipping all spins following an instantaneous change from $(\beta,h)=(1,1)$ to $(1,-1)$ (near where the evolved protocols end) is $2N$, corresponding to a value of $\phi=2N/t_0=2N/(100N)=0.02$, or about the value of the plateau immediately following the jump. The final values of entropy production identified by evolutionary learning are about an order of magnitude smaller.

The evolved low-dissipation magnetization-reversal protocols agree with those obtain by path-sampling techniques\c{rotskoff2015optimal,gingrich2016near}, showing that the way to reverse magnetization with least dissipation is to heat, reverse the field, and then cool, avoiding the large energy changes that result from the presence of finite surface tension or the large fluctuations near the critical point. The inset to panel (a) emphasizes an additional result of~\cc{gingrich2016near}, that several slightly different protocols result in similar entropy production. There may be one optimal protocol, but there are many low-lying protocols that are essentially as good (the number of nature of such protocols can be analyzed to provide insight about the structure of the trajectory ensemble\c{gingrich2016near,sriraman2005coarse}).

In panel (c) we show magnetization and protocol as a function of time, together with time-ordered snapshots, from the simulation with the lowest dissipation obtained by the learning procedure. Below that we show a protocol from a second set of evolutionary learning simulations that were instructed to find magnetization-reversal trajectories with the {\em greatest} possible entropy production. In this case the strategy is to drive the system through the phase-transition line and reverse the field only when the temperature is as low as can be achieved on the allotted timescale. The resulting dynamics is a nucleation-like mechanism in the presence of large surface tension and large thermodynamic driving force for phase change.
\begin{figure}[] 
   \centering
  \includegraphics[width=\linewidth]{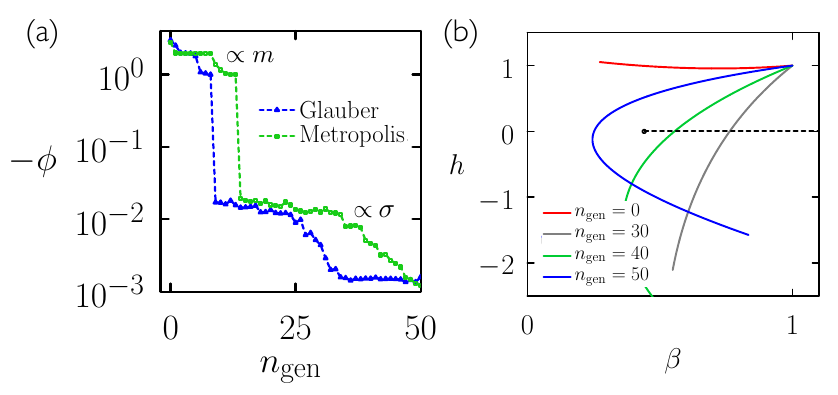} 
   \caption{Protocol learning for the Ising model is qualitatively unchanged upon going from Glauber to Metropolis dynamics. (a) Reproduction of \f{fig1}(a) (blue) together with similar results using Metropolis Monte Carlo dynamics (green). (b) Analog of \f{fig1}(b) for Metropolis dynamics. Again, the learning algorithm identifies low-dissipation trajectories as those encircling the critical point (black dot) and avoiding the first-order transition line (black dashed line).}
   \label{fig_s1}
\end{figure}

Results in this section used Ising model simulations with the Glauber acceptance probability, \eqq{glauber}. Replacing this with the Metropolis acceptance probability $p(\Delta E) = \min\left(1,\exp\left[-\beta \Delta E\right] \right))$ changes the precise dynamics of evolution of the model but not the underlying thermodynamic landscape. In \f{fig_s1} we show that evolutionary learning of low-dissipation protocols for Metropolis dynamics is qualitatively similar to that for Glauber dynamics.

\section{Low-dissipation self-assembly protocols for interacting active particles}
\label{beasties}

\begin{figure}[] 
   \centering
  \includegraphics[width=\linewidth]{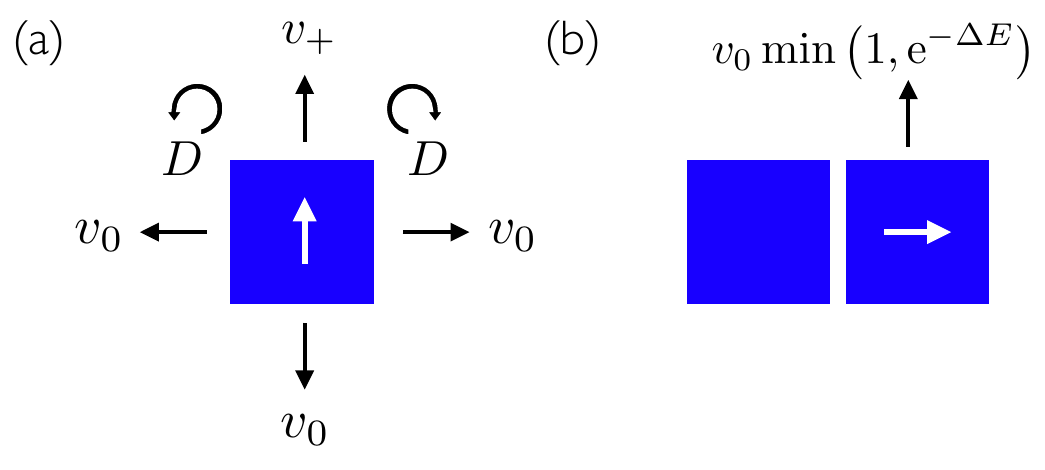} 
   \caption{Schematic of the model of active particles with interactions. (a) Rates for motion of an isolated particle. (b) Example of a rate influenced by the pairwise interparticle interaction; $\Delta E$ is the energy change of the move.}
   \label{fig2}
\end{figure}
\begin{figure*}[] 
   \centering
  \includegraphics[width=\linewidth]{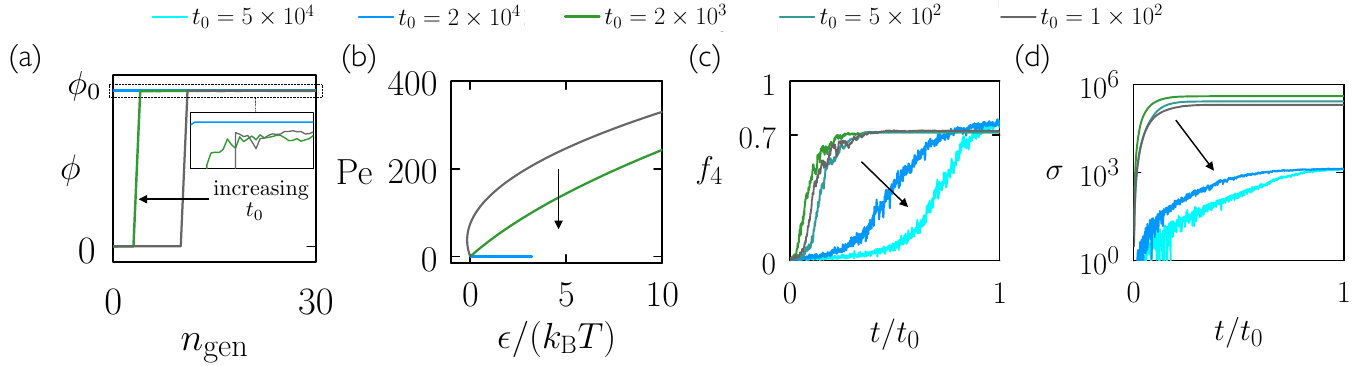} 
   \caption{Neuroevolutionary learning simulations instructed to identify a protocol $(\pp(t),\epsilon(t))$ promoting assembly with least dissipation in the model of active particles with interactions. Various values of $t_0$, the physical simulation time, are considered. (a) Order parameter \eq{phi2} as a function of evolutionary time (here and in other panels the arrow points in the direction of increasing $t_0$). (b) Parametric protocols $(\pp(t),\epsilon(t))$ learned after several evolutionary generations. (c) Solid fraction $f_4$ and (d) entropy produced as a function of simulation time, using protocols learned after several evolutionary generations.}
   \label{fig3}
\end{figure*}

Having established that the neuroevolution framework can identify protocols similar to those obtained by other methods, we now consider self-assembly in a model of active particles with interactions. The model is that of~\cc{lattice1} with pairwise nearest-neighbor interactions added, and is summarized in \f{fig2}. On a square lattice of size $N=50^2$ with periodic boundaries we consider $M=750$ particles, a packing fraction of 0.3. Particles may not overlap. Each particle bears an orientation vector (shown white in the figure) that points toward one of the 4 neighboring sites, and rotates right or left with rate $D=0.1$. Isolated particles move forward (in the direction of the orientation vector) with rate $v_+$, and in each of the other 3 directions with rate $v_0=1$. For these parameters the P\'eclet number, the dimensionless number quantifying the relative scale of self-propulsion and diffusion, is ${\rm Pe}=(v_+-v_0)/(2D)=5(v_+-1)$. Two nearest-neighbor particles possess an interaction energy $-\epsilon$, where $\epsilon>0$ indicates an attraction and $\epsilon<0$ a repulsion. Interactions modify the rates for motion via a multiplicative term $\min\left(1,\e^{-\Delta E} \right)$, $\Delta E$ being the energy change for the move. This model can self-assemble via motility-induced phase separation\c{cates2015motility}, for sufficiently large $\pp$, or via conventional phase separation, for sufficiently large $\epsilon$. Our goal is to determine which combinations of these mechanisms are required to promote low-dissipation assembly.

The system advances via a continuous-time Monte Carlo algorithm (also known as kinetic Monte Carlo or the Gillespie algorithm). If $r_i$ is the rate for process $i$ then at each step of the algorithm the process $i$ is chosen with probability $r_i/\sum_i r_i$, the sum running over all possible processes. There are $6 M$ possible processes (translations or rotations of each particle in the simulation box), with the rate for translations that result in an overlap being zero. Upon enacting the chosen process, time is advanced by an amount $-\ln \eta/\sum_i r_i$, where $\eta$ is a random number uniformly distributed on $(0,1]$. Simulations are run for time $t_0$, which we varied from of order 100 to of order $10^5$; for the chosen parameters, the characteristic time taken for an isolated particle to diffuse its own length is 1. 

At the end of the simulation we evaluate the order parameter 
\beq
\label{phi2}
\phi \equiv \begin{cases}
f_4(t_0) \qquad \qquad (f_4(t_0)\leq 0.7) \\
c_1-c_2\sigma(t_0) \quad \,\, (f_4(t_0)>0.7)
\end{cases}.
\eeq
Here $f_4$ is the fraction of particles with 4 particles as neighbors, our chosen measure of assembly. $\sigma(t_0)$ is the entropy produced over the course of the simulation. We consider the entropy production of the medium\c{seifert2005entropy}, where the entropy produced by each move is $\Delta \sigma = \ln(r/r')$. Here $r$ is the rate for the enacted process and $r'$ the rate for its reverse, which for the various move types is
\beq
\label{s1}
\Delta \sigma= \begin{cases}
0  \qquad \qquad \qquad \qquad({\rm rotation}) \\
-\beta \Delta E \qquad \qquad \quad \,\,\, ({\rm sideways}) \\
\ln v_+-\beta \Delta E \qquad \quad ({\rm forward}) \\
-\ln v_+-\beta \Delta E \qquad ({\rm backward}) 
\end{cases}.
\eeq
We define $\sigma(t_0)$ as the sum of values of \eq{s1} for each move made during the simulation. Finally, the constants $c_1=\phi_0=10^3$ and $c_2=1/(100N)$ ensure a separation of scales~\footnote{The exact numerical values of $c_1$ and $c_2$ have no effect on the outcome of learning, provided that the smallest possible value of the second clause of \eq{phi2} is always greater than the largest possible value of the first clause.} between the two clauses of \eq{phi2}, and enforce the following dual objective: the instruction to maximize \eqq{phi2} is the instruction to have the system self-assemble so that $70\%$ of its particles are in a solid-like environment, and, if so, have assembly happen with the least dissipation possible.

We use neuroevolutionary learning to find a time-dependent protocol to maximize $\phi$, as in \s{isingsec}. We start each simulation with parameters $(\pp,\epsilon)=(0,0)$, and encode the time-dependent protocol $(\pp(t),\epsilon(t))$ using a neural network. Similar to \s{ising}, the input to the neural network is the scaled time $t/t_0$, it acts 1000 times within the simulation, and its output each time it acts is $(\Delta \pp_{\bm \theta}(t),\Delta \epsilon_{\bm \theta}(t))$. After every action the simulation protocol is updated as
\bea
\epsilon(t+\Delta t) &\to&\epsilon(t)+\Delta \epsilon_{\bm \theta}(t) , \, {\rm and} \\
\pp(t+\Delta t) &\to& \max(0,\pp(t)+\Delta \pp_{\bm \theta}(t)).
\eea
The evolutionary search on the parameter set ${\bm \theta}$ proceeds as in the previous section, except that we use 50 individuals in each generation and choose the best 5 to propagate to the next generation.
 
In \f{fig3} we show the results of 5 sets of evolutionary learning simulations, each of which used a different physical time $t_0$ for the simulations (not all results are shown in each panel). In panel (a) we show the order parameter \eq{phi2} as a function of evolutionary time for simulations using three values of $t_0$; in each panel, the arrow denotes the direction of increasing $t_0$. Each $\phi$ attains a value consistent with the second clause in \eq{phi2}, showing that at least $70\%$ of particles have self-assembled into a solid-like environment. (This jump happens after fewer evolutionary steps for larger values of $t_0$, because for longer simulation times the numerical values of the control parameters $(\pp,\epsilon)$ need to be less large in order to induce assembly, and so the neural network needs fewer evolutionary generations in order to attain those values.) The largest value of $\phi$ for the largest value of $t_0$ is larger than those for the other two values of $t_0$, showing that assembly in the former case produces less entropy. We will discuss soon why this is.

In \f{fig3}(b) we show parametric plots of the protocols $(\pp(t),\epsilon(t))$ learned after several evolutionary generations (250 in the case of the two smaller times shown, 100 in the other case). For the longest time, the P\'eclet number remains zero, and only the interparticle attraction is used. In the other two cases the P\'eclet number is increased to values that promote motility-induced phase separation (a steady-state value of $\pp \gtrsim 100$ is needed to produce a solid fraction of 0.7 for packing fraction 0.3\c{lattice1}), and then the interparticle attraction is made large. Panels (c) and (d) show the outcome of these and similar protocols, showing the fraction $f_4$ of solid-like particles and the entropy $\sigma$ produced within a trajectory. The latter panel shows two sets of distinctly different entropy production rates. As we shall describe, these result from the presence or absence of self-propulsion.

 \f{fig4} confirms this behavior, showing the smallest values of entropy produced at the end of the simulation, $\sigma(t_0)$, after several generations of evolutionary learning using a range of values of $t_0$. A crossover between two different types of behavior is evident.

\begin{figure}[] 
   \centering
  \includegraphics[width=0.9\linewidth]{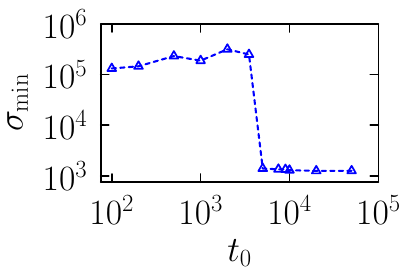} 
   \caption{Least entropy required to induce assembly, as a function of physical time $t_0$, using protocols learned after several evolutionary generations. Two distinct mechanisms are evident.}
   \label{fig4}
\end{figure}

These results reveal a change of low-dissipation assembly mechanism upon changing $t_0$. When the simulation time $t_0$ is sufficiently large, assembly of the required amount of material can proceed in the presence of diffusive motion and interparticle attractions. But diffusion alone cannot achieve assembly of the required amount of material when $t_0$ is made too small, in which case self-propulsion is required. However, propulsion is a costly mechanism, and results in considerable entropy production. 

To see this, and to understand the scale of \f{fig4}, consider that interaction-driven diffusive assembly requires the gain of about $f Nz/2$ interparticle bonds, where $f=0.7$ is the required solid-like fraction, $N=750$ is the number of particles, and $z=4$ is the coordination number. Assuming that these bonds are formed in the presence of energy scales $\epsilon$ of order 1 (in units such that $\kt=1$), the total entropy produced (the sum of bond-energy changes) is about $10^3$, consistent with the location of the lower plateau in \f{fig4}. By contrast, self-propelled particles produce entropy of order $\ln \pp$ per step. If $N=750$ particles each swim of order the box length $L = 50$ at $\pp=100$ they would produce an amount of entropy of order $10^5$, consistent with the scale of the higher plateau in \f{fig4}.

\begin{figure}[] 
   \centering
  \includegraphics[width=\linewidth]{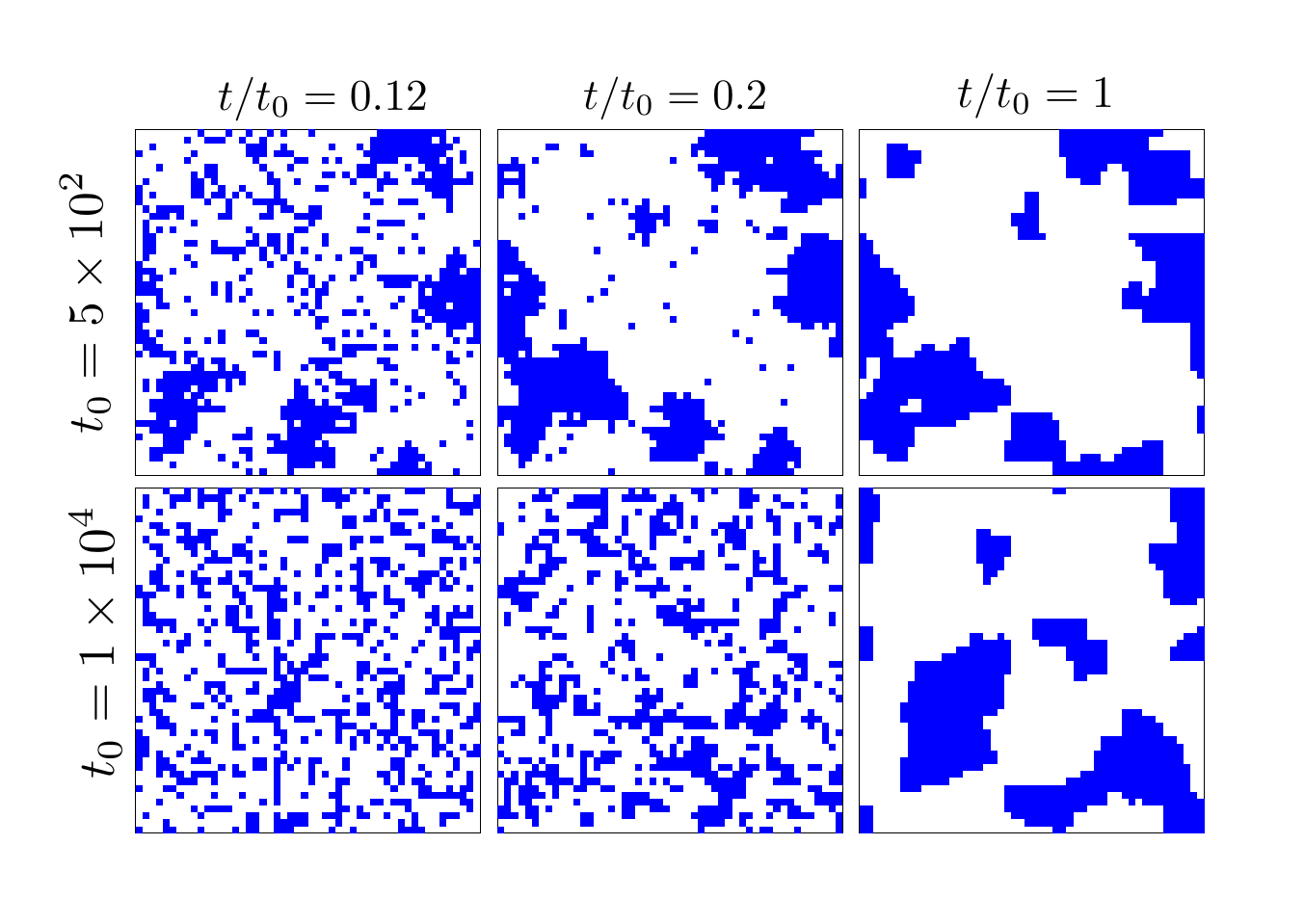} 
   \caption{Time-ordered snapshots of low-dissipation assembly involving self-propulsion (top) and diffusion only (bottom).}
   \label{fig_pic}
\end{figure}

We expect such a tradeoff to be present in general. The entropy production for energy-driven diffusive assembly scales as $N$, the number of particles, while that for motility-induced assembly scales as $N \times \ell$, the latter factor being the characteristic particle swim length. Combined with the fact that the second mechanism is faster than the first, protocols that achieve low-dissipation assembly will tend to favor diffusion when time is plentiful, and self-propulsion when it is not. Note that we have not considered the cost to change the protocol, which, for specific physical systems, may enhance or oppose this crossover.

The time at which the crossover occurs will be a function of particle concentration, which sets the characteristic interparticle separation. The timescale for diffusive encounters scales as the square of this separation, while that for encounters mediated by self-propulsion scales linearly. As a result, we expect the crossover shown in \f{fig4} to shift to larger values of $t_0$ as particle concentration is reduced. 

Time-ordered snapshots of examples of low-dissipation assembly are shown in \f{fig_pic}. There is no strong visual indication of the large difference in entropy production between these examples (self-propulsion would only be evident in a movie), but the example involving self-propulsion (top) assembles considerably faster than that involving only diffusion (bottom).

In \f{fig6} we show the outcome of three sets of evolutionary learning simulations under different constraints. All are done for simulation time $t_0=2000$. In panel (a) the neural network can control both $\pp(t)$ and $\epsilon(t)$, but in panels (b) and (c) it can control only $\pp(t)$ or $\epsilon(t)$, respectively. The strategy in (a) is to drive $\pp$ large in order to cause motility-induced phase separation, and then drive $\epsilon$ large in order to freeze the system and prevent unnecessary motion. In panel (b) the freezing mechanism is unavailable, and so the increase of $\pp$ is delayed as long as possible. Freezing is an efficient strategy: in the inset to (b) we show that case (b) produces more entropy than case (a). In case (c), diffusive motion cannot achieve the required amount of assembly.

Some protocols, particularly near the crossover in \f{fig4}, achieve assembly using moderate values of $\epsilon$ and small values of $\pp$, of order 10, not large enough to induce motility-induced phase separation but enough to speed mass transport and enable assembly that could not have happened via diffusion alone. However, because entropy production scales linearly with time but only logarithmically with $\pp$, we find that the more usual strategy is to quickly drive $\pp$ large, above 100, in order to cause motility-induced phase separation, then drive $\epsilon$ large in order to freeze the system and stop excess entropy being produced. 

\begin{figure}[] 
   \centering
  \includegraphics[width=0.8\linewidth]{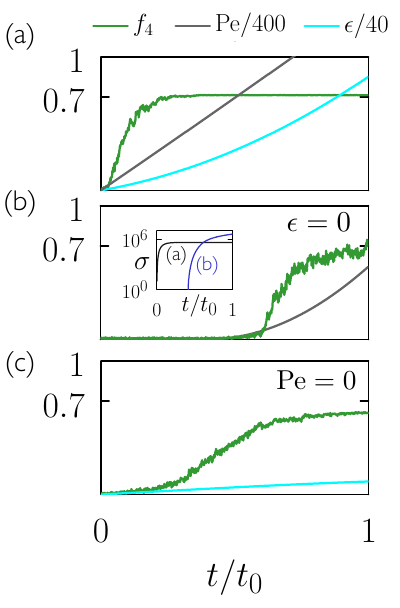} 
   \caption{Constrained assembly protocols identified by neuroevolutionary learning. (a) A low-dissipation protocol in which a neural network can control both $\pp(t)$ and $\epsilon(t)$. (b) A low-dissipation protocol in which the neural network controls only $\pp(t)$; this produces more entropy than case (a) (inset). (c) When the neural network can control only $\epsilon(t)$, the required amount of material cannot assemble in the allotted time.}
   \label{fig6}
\end{figure}

\section{Conclusions}
\label{conc}

Motivated by the competition between speed and energy efficiency in the context of biological self-assembly, we have explored self-assembly in a generic model of active particles with interactions. We have used neuroevolutionary learning to find protocols that achieve assembly within a particular time limit, and that produce as little entropy as possible. Enhanced-sampling methods such as transition-path sampling sampling\c{bolhuis2002transition} and other forms of protocol learning such as gradient-based reinforcement learning approaches could be applied to this problem\c{kaelbling1996reinforcement,sutton2018reinforcement}. In the context of reinforcement learning, evolutionary methods are relatively simple to implement (they do not require gradient computation through time) and are natural when asked to maximize an order parameter that is specified only at the final time point of a trajectory, a `sparse-reward problem' in the language of reinforcement learning (evolutionary methods can also be used with larger numbers of parameters than we have considered here\c{Guber}).

We have found that when time is plentiful, protocols use only attractive interactions and diffusive motion. When time is scarce, self-propulsion is required in order to allow the required degree of assembly to happen, but propulsion is expensive and results in more entropy being produced than in the diffusive case. 

This competition is likely to exist generically, because it depends only on simple scaling arguments and not on specific molecular details. These results, interpreted in the context of cellular assembly, suggest an incentive to maintain the system size small enough for diffusion-driven assembly to occur on the required timescales. In specific systems the question of how costly it is to impose the required protocols is also relevant, and these costs may enhance or oppose the competition identified.

\section{Acknowledgments}

This work was done as part of a User project at the Molecular Foundry at Lawrence Berkeley National Laboratory, supported by the Office of Science, Office of Basic Energy Sciences, of the U.S. Department of Energy under Contract No. DE-AC02--05CH11231. JDS acknowledges support from NIH grant R01GM141235.

%

\end{document}